\documentclass[prd,aps]{revtex4}

\usepackage{epsfig}

\newcommand{\beq}{\begin{equation}}
\newcommand{\eeq}{\end{equation}}
\newcommand{\be}{\begin{eqnarray}}
\newcommand{\ee}{\end{eqnarray}}

\begin{document}


\preprint{
\begin{tabular}{l}
\hbox to\hsize{\hfill KAIST-TH 2004/15}\\[-2mm]
\hbox to\hsize{\hfill hep-ph/yymmdd}\\[-3mm]
\hbox to\hsize{\hfill October 2004}\\[-3mm]
\end{tabular}
}

\title{
Unitarity violation of the CKM matrix
in a nonuniversal gauge interaction model
}

\author{ Kang Young Lee }
\email{kylee@muon.kaist.ac.kr}
\vskip 0.5cm

\affiliation{
Department of Physics, Korea Advanced Institute of Science and Technology,
\\
Taejon 305-701, Korea
}

\date{\today}

\begin{abstract}

We explore the unitarity violation of the CKM matrix
in the model in which the third generation fermions
are subjected to the separate SU(2)$_L$ gauge interaction.
With the recent LEP and SLC data at $Z$-pole and low-energy
neutral current interaction data, the analysis on
the parameter space of the model is updated,
and the unitary violation is predicted under the constraint.

\end{abstract}

\pacs{PACS numbers: }

\maketitle





\section{Introduction}

The Cabibbo-Kobayashi-Maskawa (CKM) matrix describes quark mixings
for the charged current interaction.  Each element of the CKM
matrix is not determined theoretically 
but just parametrized by three angles and one phase due to the
unitarity nature of the CKM matrix in the framework of 
the Standard Model (SM). 
The unitarity of the CKM matrix is a universal feature 
of the flavour physics and also holds in many new
physics scenarios like the minimal supersymmetric standard model
(MSSM) as well as in the SM.
We parametrize the unitarity relation for the first row of the matrix as 
\be
|V_{ud}|^2+ |V_{us}|^2+ |V_{ub}|^2 = 1 - \Delta, 
\ee 
and $\Delta$ should be zero in the SM. Experimentally each element 
has been measured through various weak interaction processes. 
Recently the precision test on the CKM matrix becomes possible and the
unitarity check on the first row of the CKM matrix has been
performed \cite{heidelberg}. From the recent nuclear $\beta$-decay
data, the first element is determined that $|V_{ud}| = 0.9740 \pm
0.0005$. Combined with $|V_{us}| = 0.2196 \pm 0.0023$ from kaon
decays and $|V_{ub}| = 0.0036 \pm 0.0009$ from $B$ decays, it
leads to $\Delta = 0.0031 \pm 0.0014$, which indicates a violation
of the unitarity by 2.2 $\sigma$ standard deviation \cite{abele}.
However, recent experimental results on $|V_{us}|$ suggests
unitarity again with good precision \cite{czarnecki}.

If the violation of the unitarity relation in the CKM matrix turns
out to be true, it is a clear evidence of the new physics beyond
the SM. It suggests an interesting constraint for the nature of
the possible new physics since the conventional supersymmetric
models and GUT models usually preserve the unitarity of the CKM
matrix. Examples of model with nonunitary CKM matrix contains the
model with a heavy singlet quark where the mixing between the
singlet quark and ordinary quarks can violate unitarity. 
Another example is the model with fourth generation, 
where the $4\times4$ quark mixing matrix
is unitary while $3 \times 3$ mixing matrix is not.

The nonuniversal gauge interaction is also a possible candidate
leading to the non-unitarity of the CKM matrix. The mass matrices
of up-type and down-type quarks are diagonalized by a biunitary
transformation with two independent unitary matrices. 
The charged current interactions in terms of physical states are 
expressed by the product of the two unitary transform matrices 
for left-handed up-type quarks and down-type quarks, 
which should be unitary since the mixing matrix is
the product of two unitary matrices in the SM.
However, if the SU(2) gauge interaction is nonuniversal,
the mixing matrix consists of \be V_{\rm CKM} = V_U^\dagger \left(
\begin{array}{ccc}
             a&0&0 \\
             0&b&0 \\
             0&0&c \\
           \end{array}
                 \right) V_D,
\ee which is no more unitary unless $a=b=c$. We consider a model
which contains a nonuniversal gauge interaction to predict a
non-unitary CKM matrix in this work.

The flavour physics on the third generation has drawn much interest
as a laboratory where new physics manifests.
The forward-backward asymmetry for the $Z \to b \bar{b}$ process
still shows more than 2-$\sigma$ deviation from the SM prediction
and the large value of top quark mass may imply a new dynamics.
Thus we consider the model where a separate SU(2) group
acts only on the third generation while the first two generations
couple to the usual SU(2) group \cite{malkawi1,nandi}.
This gauge group can arise as the theory
at an intermediate scale in the path of gauge
symmetry breaking of noncommuting extended technicolor
(ETC) models \cite{ETC}, in which the gauge groups for
extended technicolor and for the weak interactions
do not commute.
Due to the nonuniversality of the gauge couplings,
the CKM matrix is not unitary in this model,
although the unitarity violation is suppressed by the heavy scale
of new physics.
Moreover the flavour-changing neutral current (FCNC) interactions
generically emerge in this model, since the neutral currents are
no more simultaneously diagonalized with the corresponding mass matrix. 
Together with a new spectrum of gauge bosons, the FCNC interactions
predict various new phenomena at colliders and low-energy experiments,
and highly constrain the model parameters.
The phenomenology of this model has been intensively studied 
in the literatures, using the $Z$-pole data 
\cite{malkawi1,lee1,nandi,malkawi2} and
the low-energy data \cite{lee2,malkawi2}.

In this work, we estimate the unitarity violation in the proposed
model and show that the unitarity violation data can be explained
in this model constrained under the $Z \to b \bar{b}$
process at the $Z$-pole and low-energy neutral current experiments.
We update the previous analysis in Ref. \cite{lee1} 
with the recent LEP and SLC data \cite{LEPEWWG}. 
The atomic parity violation (APV), the neutrino-nucleon scattering,
$\nu e \to \nu e$ scattering and polarized $e$-$N$ scattering data 
are also considered to constrain the model parameters describing
the new physics scale and $Z$--$Z'$ mixing.
The observable quantity in the SM is the only product of
the unitary transform matrices for the left-handed up-type and
down-type quarks. However in this model, the matrix elements of
the individual unitary transform matrices manifest in some
processes involving the third generation and 
we might confront doubled parameters on the quark mixing. 
Thus we have to be careful in treating new parameters. 

This paper is organized as follows : In section 2, we briefly
review the model with a separate SU(2) symmetry for the third
generation and show the unitarity violation in this model. The
$Z$-pole data in the LEP and SLC experiments and low-energy
neutral current interaction experiments are analyzed to constrain
the parameter space of the model in section 3. The unitarity
violating term is predicted under the constraints in section 4 and 
we conclude in section 5.

\section{The Model}

We consider the model
based on the extended electroweak gauge group
$SU(2)_l \times SU(2)_h \times U(1)_Y $.
The first and the second generations couple to $SU(2)_l$ group
and the third generation couples to $SU(2)_h$ group.
We assign that the left--handed quarks and leptons
of the first and second generations
transform as (2,1,1/3), (2,1,-1)
and those in the third generation as (1,2,1/3), (1,2,-1)
under $SU(2)_l \times SU(2)_h \times U(1)_Y $,
while right--handed quarks and leptons transform as (1,1,2$Q$)
with the electric charge $ Q = T_{3l} + T_{3h} + Y/2 $.
We write the covariant derivative as
\begin{equation}
 D^{\mu} = \partial^{\mu} + i g_l T^{a}_{l} W^{\mu}_{l,a}
          + i g_h T^{a}_{h} W^{\mu}_{h,a}
          + i g^{\prime} \frac{Y}{2} B^{\mu} ,
\end{equation}
where $ T^{a}_{l,h}$ denotes the $SU(2)_{(l,h)}$ generators and
$Y$ the $U(1)$ hypercharge. Corresponding gauge bosons are $
W^{\mu}_{l,a} , W^{\mu}_{h,a} $ and $B^{\mu}$ with the coupling
constants $g_{l}, g_{h}$ and $g^{\prime}$ respectively. The gauge
couplings are parametrized by
\begin{equation}
g_l = \frac{e}{\sin \theta \cos \phi} ,\mbox{~~~~}
g_h = \frac{e}{\sin \theta \sin \phi},\mbox{~~~~}
g^{\prime} = \frac{e}{\cos \theta}
\end{equation}
in terms of the weak mixing angle $\theta $ and the new mixing
angle $\phi$ between $SU(2)_l$ and $SU(2)_h$.

The gauge group $ SU(2)_l \times SU(2)_h \times U(1) $
breaks down into the $ SU(2)_{l+h} \times U(1)_Y $
by the vacuum expectation values (VEV)
$ \langle \Sigma \rangle = \left( \begin{array}{cc}
             u&0 \\
             0&u
           \end{array} \right)$
of the scalar field $\Sigma$,
which is the bidoublet scalar field transforming as (2,2,0)
under $ SU(2)_l \times SU(2)_h \times U(1) $.
Sequentially the gauge group breaks down into $U(1)_{em}$
at the electroweak scale $v$ 
by the VEV of the (2,1,1) scalar field $\Phi$.
We require that the first symmetry breaking scale is
higher than the electroweak scale, parametrized by
$ v^2/u^2 \equiv \lambda \ll 1 $.
We demand that both $SU(2)$ interactions are perturbative
so that the value of the mixing angle $\sin \phi$
is constrained $g_{(l,h)}^2/4 \pi < 1$,
which results in $ 0.03 < \sin^2 \phi < 0.96 $.
After the symmetry breaking,
the masses of the heavy gauge bosons are given by
\be
m_{_{W'^{\pm}}}^2 = m_{_{Z'}}^2
= \frac{m_0^2}{\lambda \sin^2 \phi \cos^2 \phi} ,
\ee
while the ordinary gauge boson masses given by
$ m_{_{W^{\pm}}}^2
= m_0^2 ( 1- \lambda \sin^4 \phi )
= m_{_{Z}}^2 \cos^2 \theta$
where $m_0 = ev/(2 \sin \theta)$.

Since the couplings to the gauge bosons for the third generations
are different from those of the first and second generations,
we can separate the nonuniversal part
from the universal part.
First we consider the charged current:
\begin{eqnarray}
 {\cal L}^{cc} &=&  {\cal L}^{cc}_{I} + {\cal L}^{cc}_{3}~~,
\ee
where $ {\cal L}^{cc}_{I} $ denotes the universal part
and $ {\cal L}^{cc}_{3} $ the nonuniversal part.
We consider the unitary matrices $V_{U}$ and $V_{D}$
diagonalizing $U$--type and $D$--type quark mass matrices
respectively.
The universal part is given by
\be
 {\cal L}_{I} &=&  {\bar U}_L {\gamma}_{\mu}
       \left[
               G_{L} W^{\mu}  + G^{\prime }_{L} W^{\prime \mu}
       \right]
         (V_U^{\dagger} V_D) D_L
       + \mbox{H.c.}
\ee
where
\begin{eqnarray}
G_L  &=&  -\frac{g}{\sqrt{2}}
              \left( 1 - \lambda \sin^4 \phi \right) I
\nonumber \\
G^{\prime}_L  &=& \frac{g}{\sqrt{2}}
              \left( \tan \phi + \lambda \sin^3 \phi \cos \phi \right) I
\nonumber \\
\end{eqnarray}
with the 3$\times$3 identity matrix $I$
and $U = (u,c,t)^{T}$ and $D = (d,s,b)^{T}$.
We define the unitary matrix
$V_{_{CKM}}^{0} \equiv V_U^{\dagger} V_D$
corresponding to the CKM matrix of the SM.
The nonuniversal part is written in terms of
mass eigenstates as:
\be
{\cal L}_3^{cc} &=&
({V^U_{31}}^{*} \bar{u}_L + {V^U_{32}}^{*} \bar{c}_L 
                           + {V^U_{33}}^{*} \bar{t}_L)
\nonumber \\
&& \times \gamma^{\mu} (X_L W^{+}_{\mu} +X'_L W^{\prime +}_{\mu} )
(V^D_{31} d_L + V^D_{32} s_L + V^D_{33} b_L)~,
\ee
where
\begin{eqnarray}
X_L  &=&  -\frac{g}{\sqrt{2}}
              \lambda \sin^2 \phi \cdot
\nonumber \\
X^{\prime}_L  &=& -\frac{g}{\sqrt{2}}
              \left( \frac{1}{\sin \phi \cos \phi}
              \right)~.
\nonumber \\
\end{eqnarray}
Because of the existence of ${\cal L}_3$, the quark mixing matrix is
no more unitary.
Moreover 
the elements of individual unitary transform matrices
$V_U$ and $V_D$ :$\{ V_{U3j},V_{D3k}\}$ manifest 
in the unitarity violating term in general,
which are not observable in the SM.
With these elements, the number of parameters increases drastically.
This property might make it be very interesting to study
lots of new phenomena in this model,
e.g. exotic CP violation or FCNC.
However, as will be shown later,
the additional parameters $\{ V_{U3j},V_{D3k}\}$
do not play a role in the present analysis.

We can read out the expressions for the neutral current
interaction terms from the interaction lagrangian.
The nonuniversality of the gauge couplings brings forth 
the FCNC interactions at tree level which
does not exist in the SM. 
The FCNC effects on the lepton number violating processes
have been discussed in Ref. \cite{lee1}. 
For the quark sector, the FCNC interaction terms evolve
with the basis used above as follows 
\be 
{\cal L}_3^{nc} &=& ({V^D_{31}}^* \bar{d}_L + {V^D_{32}}^* \bar{s}_L 
                     + {V^D_{33}}^* \bar{b}_L)~,
\nonumber \\
&& \times \gamma^{\mu} (Y_L Z^{0}_{\mu} +Y'_L Z^{\prime 0}_{\mu} )
(V^D_{31} d_L + V^D_{32} s_L + V^D_{33} b_L)~,
\ee
where
\be
Y_L &=& \frac{g}{2 \cos \theta} \cdot \lambda \sin^2 \phi
\nonumber \\
Y'_L &=& \frac{g}{2} \cdot \frac{1}{ \sin \phi \cos \phi}~,
\ee
which yield an additional contribution to the
$B - \bar{B}$ mixings and rare decay processes through $Z$ and $Z'$
exchange diagrams, and have been discussed in Ref. \cite{lee2}.
In this paper, we assume no lepton flavour violation
to concentrate on the quark mixings.

\section{Experimental constraints}

\subsection{$Z$-pole data}

We write the most general amplitude for $Z \to b \bar{b}$ decay as
\be
M(Z \to b \bar{b}) &=& \frac{g}{2 \cos \theta_W}
     \epsilon^\mu \bar{u} \gamma_\mu (g_{bV} - g_{bA} \gamma_5) u~,
\nonumber \\
&=& \frac{g}{2 \cos \theta_W}
     \epsilon^\mu \bar{u} \big( \gamma_\mu (g_{bV}^0 - g_{bA}^0 \gamma_5)
    + \Delta^L_b \gamma_\mu P_L + \Delta^R_b \gamma_\mu P_R \big) u~,
\ee
where $\epsilon^\mu$ is the polarization vector for $Z$ boson,
$g_{bV}$ ($g_{bA}$) is the vector (axial--vector) coupling of
$b$ quark pair to $Z$ boson,
and $g_{bV}^0$ ($g_{bA}^0$) is the SM tree level coupling.
In the SM, $\Delta_b^L$ and $\Delta_b^R$ are the electroweak
corrections including the dependences on $m_t$ and $m_H$.
The large mass of top quark gives rise to additional contribution
of $m_t$--dependent corrections to $\Delta_b^L$ such as
$ \Delta^{L}_b(m_t^2) |_{\rm SM} \propto m_t^2/m_{_W}^2$
\cite{akhundov}
which arises from the top quark exchange diagram.
It is absent in the cases of light quarks,
while the contribution from these diagrams to $\Delta^R_b$
is suppressed by the factor of $m_b^2/m_t^2$.

Precision measurement of the $Z$--pole data
at LEP and SLC has provided highly accurate tests
on the SM \cite{LEPEWWG,sm2}.
Still more than $-2 \sigma$ deviation of $A_{FB}^b$,
the forward--backward asymmetry of $Z \to b \bar{b}$ decay
from the SM prediction suggests a hint of the departure from the SM
in the list of the LEP and SLC data.
It is quite helpful to introduce the precision variables
for the study of the new physics effects on the electroweak data
because the new physics contributions
to the $Z$-pole observables are expected
to be comparable with the loop contributions of the SM.
Here, we use the nonstandard electroweak precision test in terms of the
$\epsilon$ variables introduced by Altarelli {\it et al.} \cite{altarelli} 
which is useful for the analysis of the $Z$-pole data
This $\epsilon$ analysis
provides a model independent way to analyze the electroweak precision data,
where the electroweak radiative corrections
containing whole $m_t$ and $m_H$ dependencies are parametrized
into the parameters $\epsilon$'s.
Thus the $\epsilon$'s can be extracted from the data
without specifying $m_t$ and $m_H$.

The universal correction terms to the electroweak form factors
$\Delta \rho$ and $\Delta k$ are defined from the vector and
axial vector couplings of lepton pairs to $Z$ boson
given by \cite{altarelli}:
\be
g_A &=& -\frac{1}{2} \left( 1+\frac{1}{2} \Delta \rho \right),
\nonumber \\
x &\equiv& \frac{g_V}{g_A} = 1 -  \sin^2 \theta^l_{\rm eff}
    = 1 - s_0^2 (1+ \Delta k),
\ee
where $s_0^2$ is the Weinberg angle at tree level, satisfying
$s_0^2 c_0^2 = \pi \alpha(m_{_Z}) /\sqrt{2} G_F m_{_Z}^2$
and $ \sin^2 \theta^l_{\rm eff}$ is the effective Weinberg angle
including the SM loop corrections to the lepton sector.
Thus we present the amplitude in Eq. (13) in terms of
the vector and axial-vector couplings
instead of the left- and right-handed couplings.
Meanwhile the leading SM correction is left-handed
and so we present the correction terms in the left-
and right-handed basis to extract the anomalous corrections solely.
Another correction term $\Delta r_W$ is obtained
from the mass ratio $m_{_W}/m_{_Z}$ by the Eq. (1)
of Ref. \cite{altarelli2}.
The parameters $(\epsilon_1, \epsilon_2, \epsilon_3)$ are defined
by the linear combinations of correction terms
as given by \cite{altarelli}
\be
\epsilon_1~&=&~\Delta \rho, \nonumber \\
\epsilon_2~&=&~c^2_0 \Delta \rho~+~\frac{s^2_0 \Delta r_W}{c^2_0-s^2_0}
               ~-~2s^2_0 \Delta k, \nonumber\\
\epsilon_3~&=&~c^2_0 \Delta \rho~+~(c^2_0-s^2_0) \Delta k,
\ee
which avoid the new physics effects being masked
by the large $m_t^2$ corrections in $\epsilon_2$ and $\epsilon_3$.
We note that $\Delta r_W$ is irrelevant for our analysis and
affects only on $\epsilon_2$.
Hence we lay aside $\epsilon_2$ in this paper.

The parameter $\epsilon_b$ is introduced to measure
the additional contribution to the $Z b \bar{b}$ vertex
due to the large $m_t$-dependent corrections in the SM.
Since the leading contribution of the electroweak radiative correction
of the SM given in Eq. (13) is left-handed in the large $m_t$ limit,
Altarelli {\it et al}. have defined $\epsilon_b$
through the effective couplings $g_{bA}$ and $g_{bV}$
in the following manner \cite{altarelli2}:
\be
g_{bA} &=& -\frac{1}{2} \left( 1+\frac{1}{2} \Delta \rho \right)
                             (1+\epsilon_b)~,
\nonumber \\
x_b &\equiv& \frac{g_{bV}}{g_{bA}} =
\frac{ 1-\frac{4}{3} s_0^2 (1+ \Delta k) + \epsilon_b}
     {1+\epsilon_b}~.
\ee
of which asymptotic contribution is given by
$\epsilon_b \approx -G_F m_t^2/4 \pi^2 \sqrt{2}$.
The parameter $\epsilon_b$ defined in this expression
is identical to $-\Delta_b^{SM}(m_t^2)$ given in Eq. (14).

Since $\epsilon_b$ cannot be the most general expression 
for $Z b \bar{b}$ vertex, we have to introduce a new parameter 
to describe the additional right--handed current interaction effects 
as done in Ref. \cite{afb,chaylee}.
We define the correction terms $\Delta \rho_b$ and 
$\Delta k_b$ in an analogous way to those of lepton sector:
\be
g_{bA} &=& g_{A} (1+\Delta \rho_b)
       = -\frac{1}{2} \left( 1+\frac{1}{2} \Delta \rho \right) 
          (1+\Delta \rho_b) ,
\nonumber \\
\sin^2 \theta^b_{\rm eff} &=& \sin^2 \theta^l_{\rm eff} (1+\Delta k_b)
= s_0^2 (1+\Delta k) (1+\Delta k_b)~,
\ee
where $\Delta \rho_b$ is a deviation of $g_{bA}$
from the axial coupling of lepton sector $g_A$,
and $\Delta k_b$ is introduced through the effective Weinberg angle 
for $b$ quark sector.
To avoid being masked by the electroweak radiative corrections 
of the SM, we define the new epsilon parameters by the relations
\be
\epsilon_b \equiv \Delta \rho_b,
~~~~~~
\epsilon'_b \equiv \frac{2}{3} s_0^2 (\Delta \rho_b + \Delta k_b),
\ee
with canceling $\Delta^{SM}_b(m_t^2)$ in $\epsilon'_b$.
Note that we have $ \Delta \rho_b =-\Delta k_b = -\Delta_b^{SM}(m_t^2) $
in the SM, $\epsilon_b$ goes to the original definition in Eq. (16) 
and $\epsilon'_b = 0$.
Consequently $\epsilon'_b$ purely measures the
anomalous right-handed current interaction.

The parameters $\epsilon_1$ and $\epsilon_3$ are extracted
from the inclusive partial decay width $\Gamma_l$
and the forward-backward asymmetry $A_{FB}^l$
and $\epsilon_b$ and $\epsilon'_b$ are extracted
from the observables of the inclusive decay width $\Gamma_b$
and forward-backward asymmetry of $b \bar{b}$ production $A_{FB}^b$.
In consequence, the four parameters
$\epsilon_1$, $\epsilon_3$, $\epsilon_b$, and $\epsilon'_b$
are set to be one to one correspondent to the observables
$\Gamma_l$, $A_{FB}^l$, $\Gamma_b$, and $A_{FB}^b$.
The quadratic $m_t$ dependences of the SM electroweak radiative
correction appears in $\epsilon_1$ and $\epsilon_b$
while the $m_t$ dependence of $\epsilon_3$ is logarithmic.

According to the definitions,
we can express the observables in terms of $\epsilon_i$'s
up to the linear order as given in Eq. (123) of Ref. \cite{sm2},
which make it easy to perform the numerical analysis.
We give the modification of the linearized relations
by excluding the equation of $m_W^2/m_Z^2$
and adding the modified equations
between the observables $\Gamma_b$, $A_{FB}^b$
and the $\epsilon$ parameters as
\be
\Gamma_b &=& \Gamma_b \vert_B (1 + 1.42 \epsilon_1 - 0.54 \epsilon_3
+2.29 \epsilon_b)
\nonumber \\
A_{FB}^b &=& A_{FB}^b \vert_B (1 + 17.5 \epsilon_1 - 22.75 \epsilon_3
+0.157 \epsilon_b)
\ee
while the relations of $\Gamma_l$ and $A_{FB}^l$
remains intact
\be
\Gamma_l~&=&~\Gamma_l\vert_B (1+ 1.20\epsilon_1 - 0.26\epsilon_3),
\nonumber \\
A^l_{FB}~&=&~A^l_{FB} \vert_B (1+ 34.72\epsilon_1 - 45.15\epsilon_3),
\nonumber
\ee
with $s_0^2 = 0.2311$.
The Born approximation values
$\Gamma_b \vert_B$ and $ A_{FB}^b \vert_B $ are
defined by the tree level
results including pure QED and pure QCD corrections and
consequently depend upon the values of $\alpha_s(m_{_Z}^2)$
and $\alpha(m_{_Z}^2)$.
The QCD corrections to the forward-backward asymmetries
of $b$ quark pair in $Z$ decays are given
in the Ref. \cite{abbaneo,ravindran}.
We obtain
$ \Gamma_b |_B = 379.8$ MeV, and $A_{FB}^b |_B = 0.1032$
with the values $\alpha_s(m_{_Z}^2) = 0.119$
and $\alpha(m_{_Z}^2) =1/128.90$.
For the experimental analysis of $A_{FB}^b$,
a bias factor has to be introduced to scale the QCD corrections.
Here we used the average value, $s_b=0.435$,
given in Ref. \cite{abbaneo}.
Through these four linear equations
between $(\Gamma_l, A_{FB}^l, \Gamma_b, A_{FB}^b)$
and $(\epsilon_1,\epsilon_3,\epsilon_b,\epsilon_b')$,
we obtain the 4-dimensional ellipsoid in
$(\epsilon_1,\epsilon_3,\epsilon_b,\epsilon_b')$ space
from the recent LEP$+$SLC data given in Table I,
which yields
\be
\epsilon_1 &=& (5.1 \pm 1.1) \times 10^{-3},
\nonumber \\
\epsilon_3 &=& (3.8 \pm 1.8) \times 10^{-3},
\nonumber \\
\epsilon_b &=& (2.8 \pm 2.9) \times 10^{-2},
\nonumber \\
\epsilon'_b &=& (3.5 \pm 4.0) \times 10^{-2}.
\ee

We write the vector and axial vector couplings of fermions
to $Z$ boson at tree level as
\be
g_V &=& T_{3h}+T_{3l} - 2 Q \sin^2 \theta_W
        + \lambda \sin^2 \phi (T_{3h} \cos^2 \phi - T_{3l} \sin^2 \phi),
\nonumber \\
g_A &=& T_{3h}+T_{3l}
        + \lambda \sin^2 \phi (T_{3h} \cos^2 \phi - T_{3l} \sin^2 \phi),
\ee
where $\sin^2 \theta_W$ is defined as the shifted Weinberg angle
at tree level corrected by the new contribution
\be
\sin ^2 \theta_W = s_0^2 - \lambda^4 \phi \frac{c_0^2 s_0^2}{c_0^2-s_0^2},
\ee
up to the linear order of $\lambda$.
With these shifted couplings, we obtain the additional
contributions to $\epsilon$ parameters in our model,
\be
\epsilon_1^{\rm new} &=& -2 \lambda \sin^4 \phi,
\nonumber \\
\epsilon_3^{\rm new} &=& - \lambda \sin^4 \phi,
\nonumber \\
\epsilon_b^{\rm new} &=&  \lambda \sin^2 \phi (1+|V^D_{33}|^2 ),
\nonumber \\
\epsilon_b^{\prime \rm new} &=&  \lambda \sin^2 \phi |V^D_{33}|^2 ,
\ee
which is expressed by three parameters, $\lambda$, $\sin \phi$
and $V^D_{33}$.

With the experimental ellipsoid given in Eq. (20),
the allowed parameter space is given by
\be
\lambda < 0.126~~(0.14)~~~~  {\rm at}~~90 \%~~ (95 \%)~~ {\rm C.L.},
\ee
and $ \sin \phi$ and $|V^D_{33}|$ are unconstrained.
This bound is corresponding to
$m_{Z'} > 1.15$ $(1.10)$ TeV..
We plot the model predictions in terms of model parameters
with the experimental ellipses for 1-$\sigma$ and 2-$\sigma$
confidence level in Fig. 1.
The left upper plot is given on the $\epsilon_1-\epsilon_3$ plane,
the right upper plot on the $\epsilon_1-\epsilon_b$ plane,
the left lower plot on the $\epsilon_1-\epsilon'_b$ plane,
the right lower plot on the $\epsilon_3-\epsilon_b$ plane.
The straight lines denote the model predictions and
the SM predictions is expressed by the black dots.
We vary the parameter $\sin^2 \phi$ and fix $|V^D_{33}|$ to be 1.
The parameter $|V^D_{33}|$ far from 1 is not likely since
$|V_{tb}| \approx 1$.

\subsection{Low energy experiment}

The low-energy neutral current interactions such
as $\nu e\rightarrow \nu e$,  $\nu N$ scattering, and $e_{L,R}N
\rightarrow e_{L,R} X$ are expressed by the effective four-fermion
interactions and the coefficients of four-fermion operators
have been precisely measured.
The SM predictions for the coefficients are obtained
including radiative corrections,
while our model predictions involve 
both $Z$ and $Z^{\prime}$ contributions at tree level.
We perform the analysis to constrain the model parameters
with the low-energy data.

For neutrino-hadron scattering, the relevant effective Hamiltonian
can be written as
\begin{equation}
H^{\nu N} = \frac{G_F}{\sqrt{2}} \overline{\nu}
\gamma^{\mu} (1-\gamma_5) \nu
\sum_i \Bigl[ \epsilon_L(i) \overline{q}_i \gamma_{\mu}
(1-\gamma_5) q_i + \epsilon_R (i) \overline{q}_i \gamma_{\mu} (1+
\gamma_5) q_i \Bigr],
\end{equation}
where $\epsilon_{L,R} (i)$ ($i=u,d$) are given by
\be
\epsilon_{L,R} (u,d) = \epsilon_{L,R}^{\rm SM} (u,d)
                       (1 - \lambda \sin^4 \phi).
\ee

The relevant effective Hamiltonian
for the scattering $\nu e \rightarrow \nu e$
at low energy can be written in the form
\begin{equation}
H^{\nu e} = \frac{G_F}{\sqrt{2}} \overline{\nu}
\gamma^{\mu}  (1-\gamma_5) \nu \overline{e} \gamma_{\mu} (g_V^{\nu e}
- g_A^{\nu e} \gamma_5) e.
\end{equation}
The coupling constants $g_V^{\nu e}$ and $g_A^{\nu e}$ are written as
\be
g_{V(A)}^{\nu e} = g_{V(A)}^{\nu e} |_{\rm SM}
                   (1 - \lambda \sin^4 \phi).
\ee

For electron-hadron scattering such as $e_{L,R} N \rightarrow eX$
performed in the SLAC polarized electron experiment, the
parity-violating Hamiltonian  can be written as
\begin{equation}
H^{eN} = -\frac{G_F}{\sqrt{2}} \sum_i \Bigl[ C_{1i} \overline{e}
\gamma^{\mu} \gamma_5 e \overline{q}_i \gamma_{\mu} q_i + C_{2i}
\overline{e} \gamma^{\mu} e \overline{q}_i \gamma_{\mu} \gamma_5 q_i
\Bigr].
\label{hen}
\end{equation}
The coefficients $C_{1i}$ are given by
\be
C_{1u,d} = C_{1u,d}^{\rm SM} (1 - \lambda \sin^4 \phi)
\ee
while the coefficients $C_{2i}$ given by
\be
C_{2u} &=& C_{2u}^{\rm SM} (1 - \lambda \sin^4 \phi)
          + 2 \lambda |V^U_{31}|^2 \sin^2 \phi \sin^2 \theta_W,
\nonumber \\
C_{2d} &=& C_{2d}^{\rm SM} (1 - \lambda \sin^4 \phi)
          - 2 \lambda |V^D_{31}|^2 \sin^2 \phi \sin^2 \theta_W.
\ee
Note that only $C_{2i}$ involve the additional parameters
$V^{U,D}_{31}$ because of the operator structure.

The experimental values for the observables and the
standard model predictions are tabulated in Table~II
referring to the particle data book \cite{pdb}.
We perform the $\chi^2$ fit for the ten physical observables 
listed in Table~II with respect to the parameters
$\lambda$, $\sin \phi$ and $|V^{U,D}_{3j}|$.
The best fit value is obtained at
\be
\lambda = 0.435
~~~~\sin \phi = 0.25,
~~~~|V^U_{33}| = 0.7,
~~~~|V^D_{33}| = 0,
~~~~ m_{Z'} = 503.76~(\rm GeV),
\ee
with $\chi^2/dof = 11.331/10$.
Under the LEP bound of Eq. (26), the best fit values are
\be
\lambda = 0.112
~~~~\sin \phi = 0.35,
~~~~|V^U_{33}| = 1,
~~~~|V^D_{33}| = 0,
~~~~ m_{Z'} = 732.99~(\rm GeV),
\ee
with $\chi^2/dof = 11.332/10$.
We note that the $\chi^2_{\rm min}$ is very close to the SM
value $\chi^2_{\rm SM}/dof = 11.35/10$ and no absolute bound
for $\lambda$ and $\sin \phi$ from the low-energy data. 

The atomic parity violation can be described by the Hamiltonian in
Eq.~(\ref{hen}). The weak charge of an atom is defined as
\begin{equation}
Q_W = -2 \Bigl[ C_{1u} (2Z+N) + C_{1d} (Z+2N) \Bigr],
\end{equation}
where $Z$ ($N$) is the number of protons (neutrons) in the atom. For
${}^{133}_{55} \mbox{Cs}$  atom, $Z=55$, $N=78$, the correction to the
weak charge is given by
\begin{eqnarray}
\Delta Q_W &\equiv& Q_W - Q_W^{\mathrm{SM}}
= -2 \Bigl[ \Delta C_{1u} (2Z+N) + \Delta C_{1d} (Z+2N) \Bigl]
\nonumber \\
&=& \Delta Q_W^{\rm SM} (1 - \lambda \sin^4 \phi).
\end{eqnarray}
The recent measurements and the analyses of the weak charge for the Cs
and Tl atoms give the value \cite{apv_exp}.
\be
Q_W  &=& -72.69 \pm 0.48 ~~~~~~ ({\rm Cs}),
\nonumber \\ 
Q_W  &=& -116.6 \pm 3.7 ~~~~~~~ ({\rm Tl}),
\ee
while the SM predictions are $ Q_W = -73.19 \pm 0.03$ (Cs),
and $ Q_W = -116.81 \pm 0.04$ (Tl).
The bound from Cs atom is stronger than that from Tl atom.
Figure 2 shows the allowed region of the parameter space 
($\lambda, \sin^2 \phi$) by the low-energy neutral current data and
the atomic parity violation data at 90 \% confidence level.
We find that the constraint of the atomic parity violation
is stronger than that of the $\chi^2$ fit of the low-energy 
neutral current data.

\section{Unitarity violation}

We have the modified CKM matrix in the lagrangian:
\be
V_{_{CKM}} = V_{_{CKM}}^0 + \left( \begin{array}{ccc}
       {V^U_{31}}^*V^D_{31} & {V^U_{31}}^*V^D_{32} & {V^U_{31}}^*V^D_{33} \\
       {V^U_{32}}^*V^D_{31} & {V^U_{32}}^*V^D_{32} & {V^U_{32}}^*V^D_{33} \\
       {V^U_{33}}^*V^D_{31} & {V^U_{33}}^*V^D_{32} & {V^U_{33}}^*V^D_{33} \\
                            \end{array}
                          \right) \cdot \lambda~\sin^2 \phi~,
\ee
which describes the quark mixings for the charged currents
coupled to the $W^{\pm}$ bosons.
The mixing matrix for $W'^{\pm}$ bosons has the same structure
as above matrix except that the model parameter $\lambda \sin^2 \phi$
is replaced by $1/\sin \phi \cos \phi$.

Although the quark mixing matrix is given in Eq. (37),
the `observed' CKM matrix measures the effective
charged current interactions
through  both $W$ and $W'$ exchange diagrams at low energies.
We consider the effective Hamiltonian for extracting $|V_{uq}|$
\be
{\cal H}_{\rm eff} = \frac{G_F}{\sqrt{2}} \sum_{q=d,s,b} V_{uq}
         ( \bar{u} \gamma_\mu (1-\gamma_5) q)
         ( \bar{\nu} \gamma^\mu (1-\gamma_5) l) + {\rm H.c.},
\ee
where the measured CKM matrix element is
$V_{uq} = V^0_{uq} (1 - \lambda \sin^4 \phi) $
and $ V^0_{uq}$ is the CKM matrix element in the SM.
The muon decay constant $G_F$ is identical to that in the SM
if no lepton mixing is assumed \cite{malkawi2}.
The additional parameters $\{ V_{U3j},V_{D3k}\}$ are canceled
in Eq. (38) when we sum the $W$ and $W'$ contributions
in the leading order of $\lambda$.
Thus the unitarity violating term defined in Eq. (1) is derived
\be
\Delta = 2 \lambda \sin^4 \phi.
\ee
We show the unitarity violation $\Delta$ with respect to the
$M_{Z'}$ in Fig. 3.
With the best fit value of Eq. (33) from the low-energy neutral current data,
under the constraints from the LEP bound and 
the atomic parity violation bound,
we obtain the unitarity violating term as
\be
\Delta \approx 0.0034,
\ee
which is close to the reported value 
$\Delta = 0.0031 \pm 0.0014$ \cite{abele}.

\section{Concluding remarks}

We explained the recent measurement on the unitarity violation
of the CKM matrix by introducing the new physics with
the separate SU(2) gauge symmetry on the third generation.
The additional SU(2) symmetry leads to the nonuniversality
of the electroweak gauge coupling, which leads to 
the unitarity violating term $\Delta$ of order ${\cal O}(v^2/u^2)$ 
in the quark mixing matrix. 
We performed the updated analysis on the model parameters
with the recent LEP+SLC data, low-energy neutral current data,
and the atomic parity violation data.
The model prediction for $\Delta$ with the best $\chi^2$ fit 
agrees very well with the value reported 
in the Ref. \cite{heidelberg,abele}.
In conclusion, it is a very interesting and instructive 
to find new physics model to explain the violation of unitarity 
in the CKM matrix even though the violation is still controversial.
We show that the nonuniversal gauge interaction model
is a good candidate for the new physics with the unitarity violation
of the CKM matrix.
The updated and precise measurement on the CKM matrix elements
will probe the unitarity in the future.

\acknowledgments
This work was supported by Korea Research Foundation Grant
(KRF-2003-050-C00003).

\def\PRD #1 #2 #3 {Phys. Rev. D {\bf#1},\ #2 (#3)}
\def\PRL #1 #2 #3 {Phys. Rev. Lett. {\bf#1},\ #2 (#3)}
\def\PLB #1 #2 #3 {Phys. Lett. B {\bf#1},\ #2 (#3)}
\def\NPB #1 #2 #3 {Nucl. Phys. B {\bf #1},\ #2 (#3)}
\def\ZPC #1 #2 #3 {Z. Phys. C {\bf#1},\ #2 (#3)}
\def\EPJ #1 #2 #3 {Euro. Phys. J. C {\bf#1},\ #2 (#3)}
\def\JHEP #1 #2 #3 {JHEP {\bf#1},\ #2 (#3)}
\def\IJMP #1 #2 #3 {Int. J. Mod. Phys. A {\bf#1},\ #2 (#3)}
\def\MPL #1 #2 #3 {Mod. Phys. Lett. A {\bf#1},\ #2 (#3)}
\def\PTP #1 #2 #3 {Prog. Theor. Phys. {\bf#1},\ #2 (#3)}
\def\PR #1 #2 #3 {Phys. Rep. {\bf#1},\ #2 (#3)}
\def\RMP #1 #2 #3 {Rev. Mod. Phys. {\bf#1},\ #2 (#3)}
\def\PRold #1 #2 #3 {Phys. Rev. {\bf#1},\ #2 (#3)}
\def\IBID #1 #2 #3 {{\it ibid.} {\bf#1},\ #2 (#3)}

\newpage

\vskip 2.0cm

\begin{table}
\label{table1}
\begin{center}
\begin{tabular}{clclc}
\hline \hline
& && Measurement&  \\
\hline
&$m_Z$ && 91.1875 $\pm$ 0.0021~GeV&  \\
&$\Gamma_l$ && 83.984 $\pm$ 0.086~MeV&   \\
&$A_{FB}^l$ && 0.0171 $\pm$ 0.0010&  \\
&$R_b$ && 0.21638 $\pm$ 0.00066&   \\
&$A_{FB}^b$ && 0.0997 $\pm$ 0.0016&  \\
\hline \hline
\end{tabular}
\caption{ Data on the precision electroweak tests, quoted in
\cite{LEPEWWG} }
\end{center}
\end{table}

\vskip 3.0cm

\begin{table}
\label{table2}
\vspace{1.0cm}
\begin{tabular}{ccc}
\hline \hline
&Experiments& SM prediction \\ \hline
$\epsilon_L (u)$ & 0.326$\pm$0.012 & 0.3460$\pm$0.0002 \\
$\epsilon_L (d)$ & $-$0.441$\pm$0.010 & $-$0.4292$\pm$0.0001 \\
$\epsilon_R (u)$ & $-0.175^{+0.013}_{-0.004}$& $-$0.1551$\pm$0.0001 \\
$\epsilon_R (d)$ & $-0.022^{+0.072}_{-0.047}$& 0.0776 \\
\hline
$g_V^{\nu e}$ & $-$0.040$\pm$0.015& $-$0.0397 $\pm$0.0003 \\
$g_A^{\nu e}$ & $-$0.507$\pm$0.014 & $-$0.5065$\pm$0.0001\\ \hline
$C_{1u}+C_{1d}$ & 0.148$\pm$0.004 & 0.1529$\pm$0.0001\\
$C_{1u}-C_{1d}$ & $-$0.597$\pm$0.061 & $-$0.5299$\pm$0.0004 \\
$C_{2u}+C_{2d}$& 0.62$\pm$0.80 & $-$0.0095\\
$C_{2u}-C_{2d}$& $-$0.07$\pm$0.12 & $-$0.0623$\pm$0.0006\\
\hline \hline
\end{tabular}
\caption{Values of low-energy neutral current experimental data
compared to the Standard Model predictions, quoted in Ref. \cite{pdb}}
\end{table}

\newpage

\begin{figure}[htb]
\begin{center}
\hbox to\textwidth{\hss\epsfig{file=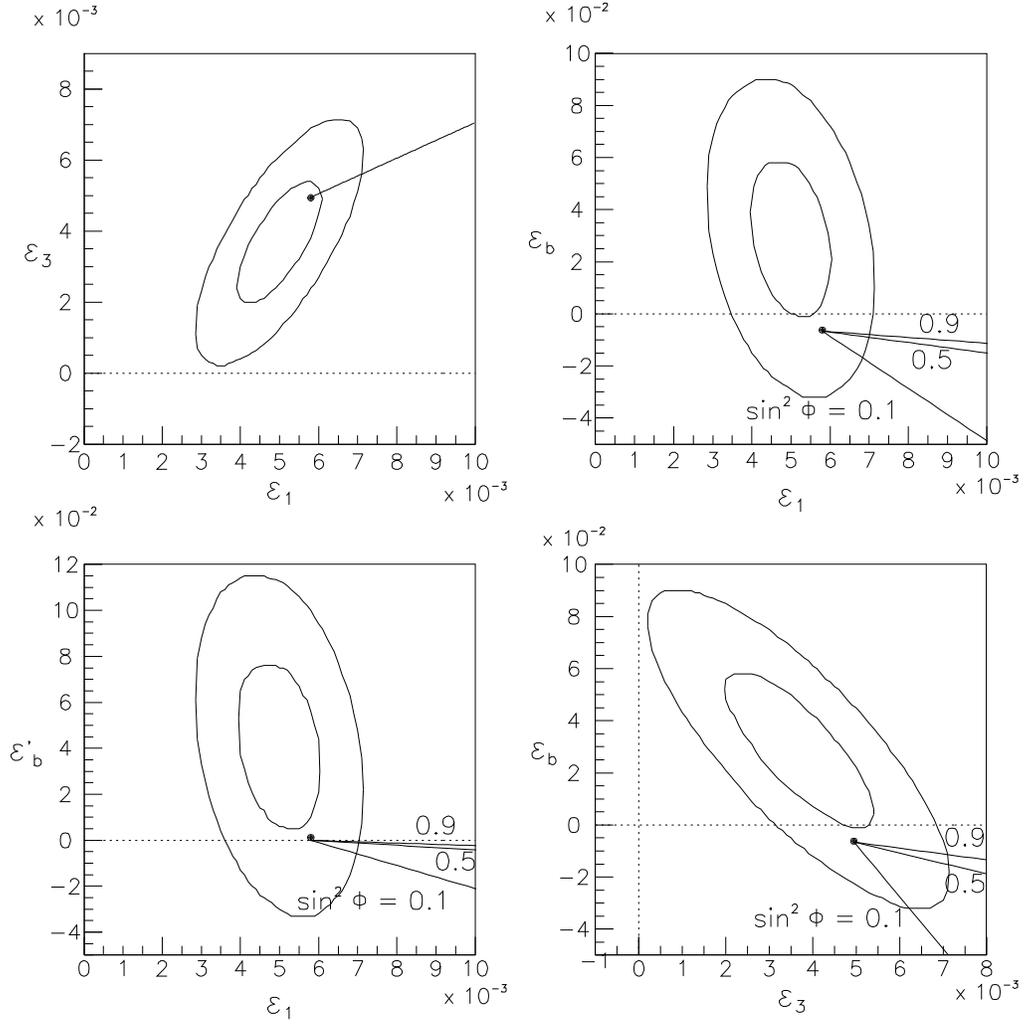,width=16cm}\hss}
\vspace{1cm}
\caption{
Model predictions and experimental ellipses in the $\epsilon_i$ space.
}
\end{center}
\end{figure}

\begin{figure}[htb]
\begin{center}
\hbox to\textwidth{\hss\epsfig{file=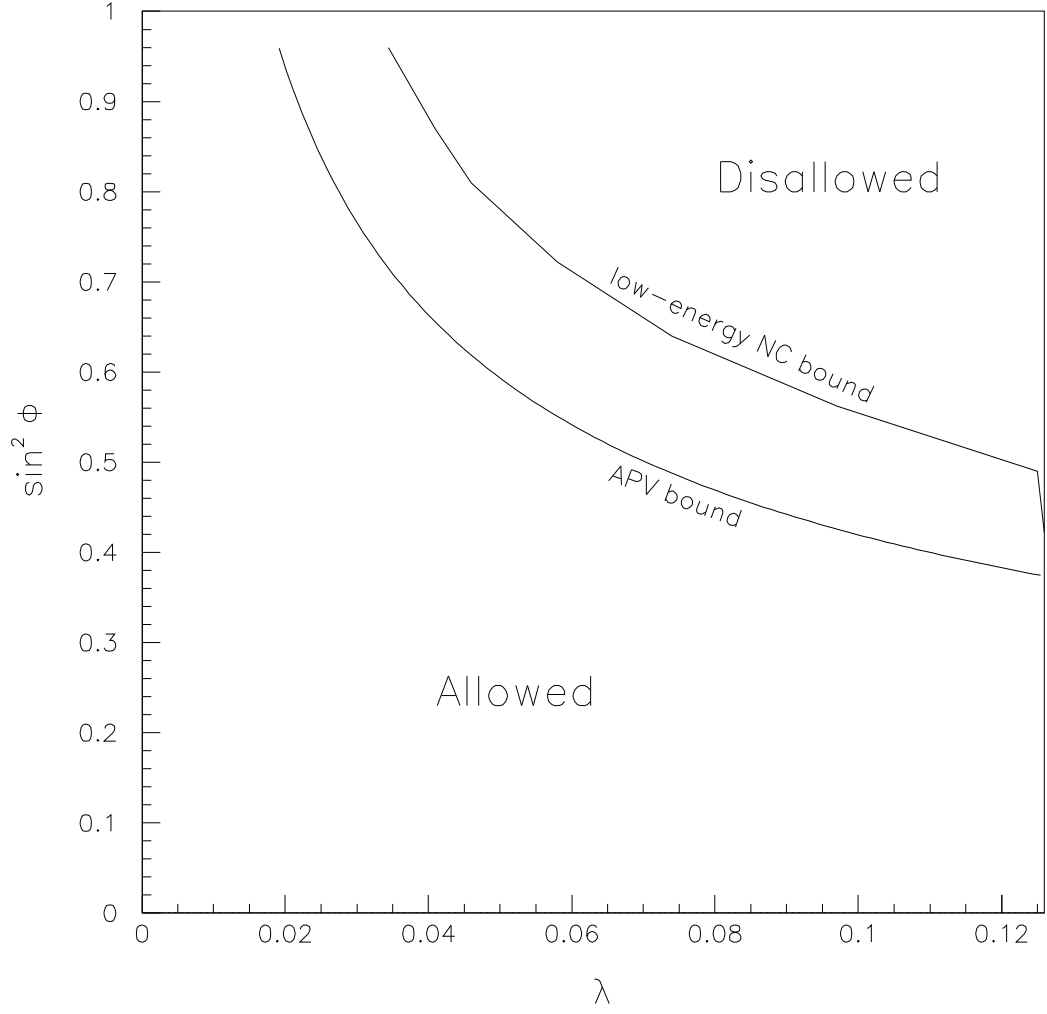,width=16cm}\hss}
\vspace{1cm}
\caption{
Allowed model parameter set on $(\lambda,\sin^2 \phi)$ plane
by the low-energy neutral current data and the atomic parity violation data
at 90 \% confidence level.
}
\end{center}
\end{figure}

\begin{figure}[htb]
\begin{center}
\hbox to\textwidth{\hss\epsfig{file=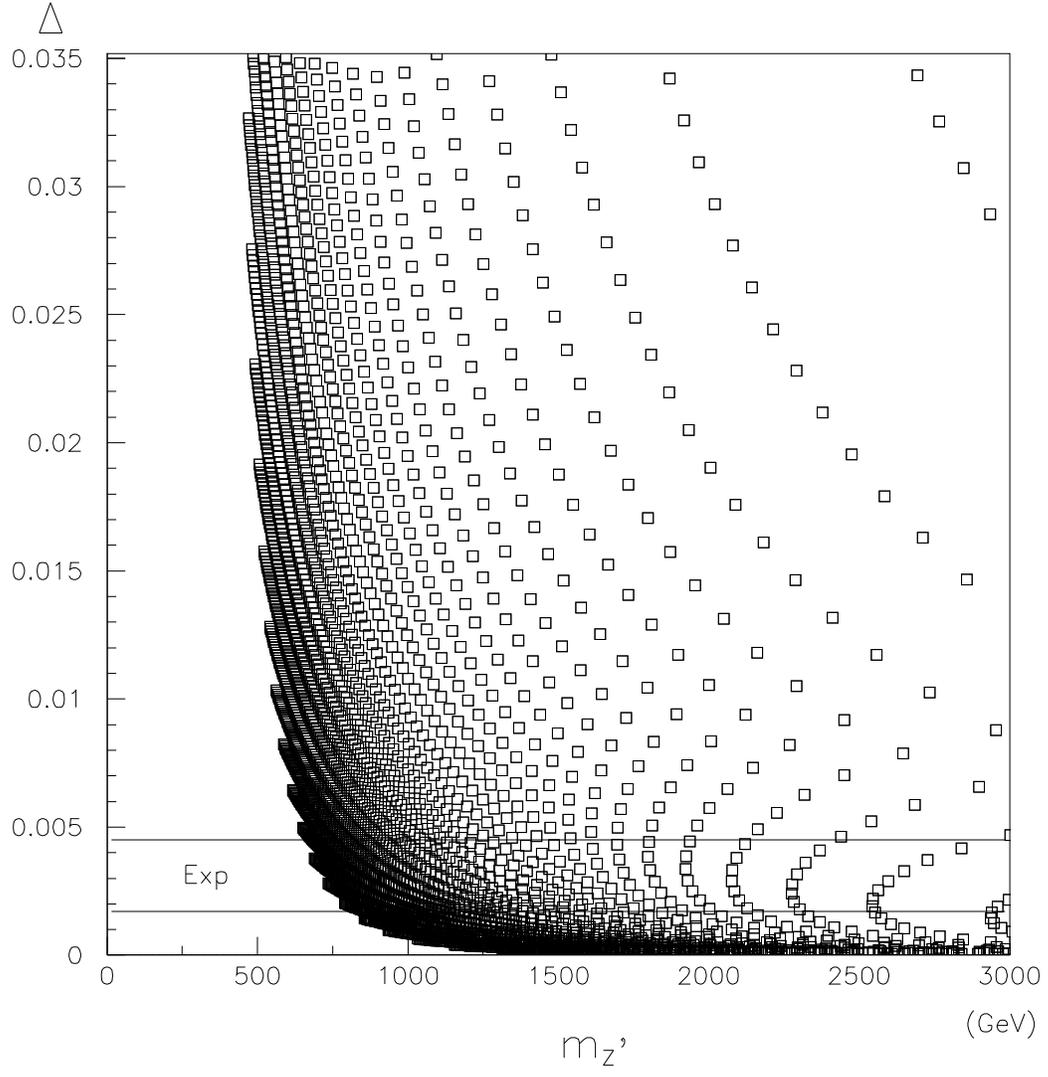,width=16cm}\hss}
\vspace{1cm}
\caption{
Model predictions for $\Delta$ with respect to $m_{Z'}$
under all constraints.
}
\end{center}
\end{figure}

\end{document}